%Paper: hep-th/9409188
%From: WITTEN@sns.ias.edu
%Date: 29 Sep 1994 14:54:33 -0400 (EDT)

\input harvmac
\newcount\figno
\figno=0
\def\fig#1#2#3{
\par\begingroup\parindent=0pt\leftskip=1cm\rightskip=1cm\parindent=0pt
\baselineskip=11pt
\global\advance\figno by 1
\midinsert
\epsfxsize=#3
\centerline{\epsfbox{#2}}
\vskip 12pt
{\bf Fig. \the\figno:} #1\par
\endinsert\endgroup\par
}
\def\figlabel#1{\xdef#1{\the\figno}}
\def\encadremath#1{\vbox{\hrule\hbox{\vrule\kern8pt\vbox{\kern8pt
\hbox{$\displaystyle #1$}\kern8pt}
\kern8pt\vrule}\hrule}}

\overfullrule=0pt

%macros
%
\def\tilde{\widetilde}
\def\bar{\overline}

\font\zfont = cmss10 %scaled \magstep1

\def\bigone{\hbox{1\kern -.23em {\rm l}}}
\def\ZZ{\hbox{\zfont Z\kern-.4emZ}}

\Title{hep-th/9409188, HUTP-94/A034, IASSNS-HEP-94-69}
{\vbox{\centerline{ON ORBIFOLDS WITH DISCRETE TORSION}}}
\smallskip
\centerline{Cumrun Vafa}
\smallskip
\centerline{\it Lyman Laboratory of Physics, Harvard University}
\centerline{\it Cambridge, MA 02138 USA}
\smallskip
\centerline{and}
\smallskip
\centerline{\it School of Natural Sciences, Institute for Advanced Study}
\centerline{\it Olden Lane, Princeton, NJ 08540, USA}
\smallskip
\centerline{and}
\smallskip
\centerline{Edward Witten}
\smallskip
\centerline{\it School of Natural Sciences, Institute for Advanced Study}
\centerline{\it Olden Lane, Princeton, NJ 08540, USA}\bigskip
\baselineskip 18pt

\medskip

\noindent
We consider the interpretation in classical geometry of conformal
field theories constructed from orbifolds with discrete torsion.
In examples we can analyze, these spacetimes contain ``stringy regions'' that
from a classical point of view are singularities that are to be neither
resolved nor blown up.  Some of these models also give particularly
simple and clear examples of mirror symmetry.
%\draft
\Date{October, 1994}

\newsec{Introduction}

Much of the fascination and mystery of string theory involves the relation
between classical and stringy geometry.  One facet of this involves
singularities of space-time; what in classical geometry is a singularity
may in string theory simply be a region in which stringy effects are large.
An elementary example involves the classical solutions of string
theory constructed as orbifolds \ref\orbi{L. Dixon, J. Harvey, C. Vafa,
and E. Witten, ``Strings On Orbifolds, I, II,''
Nucl.Phys. {\bf B261} (1985) 678,
{\bf B274} (1986) 285.}, by which we mean in this context
simply the quotient
of a torus $M$  by a finite group $\Gamma$.  If $\Gamma$ does not
act freely, the classical space-time has singularities at fixed points
of elements of $\Gamma$, but the conformal field theory of the orbifold
is nonetheless perfectly regular.

\nref\mark{D. Markushevich, M. Olshanetsky, and A. Perelomov,
``Description Of A Class Of Superstring Compactifications Related
To Semisimple Lie Algebras,'' Commun. Math. Phys. {\bf 111} (1987) 247.}
\nref\roan{S.S. Roan, ``Minimal Resolutions of Gorenstein
Orbifolds in Dimension Three,'' Acad. Sinica preprint R940606-1.}
\nref\aspin{P. Aspinwall, ``$(2,2)$ Superconformal Field Theories
Near Orbifold Points,'' Commun. Math. Phys. {\bf 128} (1990)

593, ``Resolution Of Orbifold Singularities In
String Theory,'' IASSNS-HEP-94/9.}
Particularly interesting is the case in which $M$ is a complex torus,
say of complex dimension $n$, and $\Gamma$ preserves the complex
structure of $M$ and acts trivially
on the canonical line bundle.  Then the conformal field theory
of $M/\Gamma$ has $(2,2)$ supersymmetry and integral $U(1)$ charges, just
like conformal field theories associated with smooth Calabi-Yau manifolds.
In fact, in simple cases one readily finds in the spectra of the orbifold
theory in the twisted sectors
marginal operators with the quantum numbers of elements of
$H^1(M,T)$ or $H^1(M,T^*)$, suggesting that the orbifold can
be blown up or resolved to get a smooth Calabi-Yau manifold.
This was originally noticed by hand in some simple examples \orbi;
the results have been extended in various directions

\refs{\mark,\roan,\aspin}.    When such resolution or blow-up is possible,
 the orbifold theory has the status
of a soluble special case of a (generically nonsoluble) family of
conformal field theories associated with the smooth Calabi-Yau manifold.

Orbifold conformal field theories, however, have a generalization
by turning on what is
known as discrete torsion \ref\tor{C. Vafa, ``Modular Invariance And
Discrete Torsion On Orbifolds,''
Nucl. Phys. {\bf B273} (1986) 592.}, which
involves introducing non-trivial phases to weight differently certain
path-integral sectors.  These non-trivial phases can be
introduced when $H^2(\Gamma,U(1))\not= 0$.  The issues
mentioned above have not been addressed in the context of conformal
field theory with discrete torsion; it is our intention to begin this
analysis here.
The result we will find (in the examples we will analyze) is as follows.
Addition of discrete torsion markedly changes the geometrical interpretation
of an orbifold theory; in many instances, discrete torsion gives a
theory that is not continuously connected to a theory based on a smooth
Calabi-Yau manifold.  In those cases, by resolution or deformation
one can partially eliminate the singularities, but
one remains with isolated singularities in the
classical space-time.  To a string theorist, these singularities
are simply regions in which stringy effects are large.  The discrete
torsion is supported entirely in these isolated stringy regions.

For instance, from some viewpoints
the simplest possible isolated singularity of a three-dimensional
Calabi-Yau manifold is the conifold singularity
\eqn\consing{x_1{}^2+x_2{}^2+x_3{}^2+x_4{}^2=0.}
In contrast to orbifold singularities, this type of singularity
is a singularity even in conformal field theory -- in the absence
of discrete torsion.  (For instance, Yukawa couplings
have a pole at the conifold point \ref\candelas{P. Candelas, X. della Ossa,
P. S. Green, and L. Parkes, ``A Pair Of Calabi-Yau Manifolds As An Exactly
Soluble Superconformal Field Theory,'' Nucl. Phys. {\bf B359} (1991)
21.}.)
We will find, though, that with ${\bf Z}_2$ discrete torsion, the
conifold is {\it not} a singularity in conformal field theory.

A smooth Calabi-Yau manifold (without discrete torsion) can develop
a conifold singularity through degeneration of either its complex
structure or its Kahler structure.  Both of these possibilities have
arisen in conformal field theory, and in fact \ref\whomo{D. Morrison,
``Through The Looking Glass,'' Duke preprint, to appear.} they are

apparently mirror to each other.  Conversely, the singularity
of the conifold can be removed either by deformation of complex structure,
that is, by deforming the equation to
\eqn\defqn{x_1{}^2+x_2{}^2+x_3{}^2+x_4{}^2=\epsilon,}
with $\epsilon$ a complex parameter, or by resolving the singularity.
(The relevant resolutions are the small resolutions that preserve the
Calabi-Yau condition; they are described from the point of view of
gauge theory in \ref\phases{E. Witten, ``Phases of $N=2$ Models In Two
Dimensions,'' Nucl. Phys.{\bf B403} (1993) 159. }.)
To summarize the situation, in the absence of discrete torsion there are
two known conformal field theories associated with the conifold -- one,
call it the $A$ model, in which the singularity arises by degeneration
of Kahler structure (and theta angle),\foot{The Kahler structure in the
sense of conformal field theory combines the conventional Kahler structure
with the theta angles.}
and one, call it the $B$ model, in which the singularity
arises by degeneration of complex structure.
Each model depends on
one complex parameter, which measures
the extent to which the Kahler structure ($A$ model) or complex
structure ($B$ model) has been deformed.
Such deformation is obligatory (in the $A$ model case it is enough to
deform the theta angle away from zero rather than deforming the classical
Kahler structure) since the conifold is a singularity in conformal field
theory.

In this paper, we will find a third conformal field theory of the conifold
-- call it the $C$ model.  In the $C$ model, there is a ${\bf Z}_2$
discrete torsion sitting at the  ``singularity,'' which is in fact not
a singularity in the conformal field theory sense, but just a region
in which stringy effects are essential.  The $C$ model has no marginal
operators (in particular the analogs of $H^{1,1}$ and $ H^{2,1}$ are zero),
so in the $C$ model
there is no way to modify or eliminate the ``singularity.''  This is the
opposite of the situation in the $A$ and $B$ models, in which deformation or
blowup from the singularity are obligatory.  The $A$ model has $H^{1,1}=1$,
$H^{2,1}=0$, and the $B$ model has $H^{1,1}=0$, $H^{2,1}=1$, so transitions
preserving $(2,2)$ supersymmetry cannot occur among the $A$, $B$, and $C$
models.

\subsec{Mirror Symmetry}

We have not yet mentioned another aspect of the present work which was
in fact the starting point: the connection with mirror symmetry.
In the course of analyzing simple examples of orbifolds with discrete
torsion, we will find a very simple example of mirror symmetry,
perhaps the only known type of example (apart from a complex torus)
in which mirror symmetry can be understood and demonstrated completely.

This also raises the question of what mirror symmetry does to the $C$ model.
Apparently the ``singularity'' (or more properly the stringy region)
of the $C$ model has no invariant meaning:
while conformal field theories approaching $A$ type singularities
are mirror to theories approaching $B$ type singularities, those
with $C$ type singularities can be mirror to conformal field theories
of perfectly smooth Calabi-Yau manifolds.
To put this differently, we will get an example in which a maximally
extended family of smooth Calabi-Yau manifolds is mirror to a family
of Calabi-Yau manifolds that are all singular.  This may be a much
more general phenomenon for mirror symmetry.

\newsec{A ${\bf Z}_2\times {\bf Z}_2$ Orbifold}

We need orbifolds with a finite group $\Gamma$ such that $H^2(\Gamma,U(1))$
is non-zero, so that discrete torsion is possible.  We will take
$\Gamma={\bf Z}_k\times {\bf Z}_k$; in fact, in our examples, $k$ will
mainly be 2 or 3.

One has $H^2({\bf Z}_k\times {\bf Z}_k,U(1))\cong {\bf Z}_k$, so discrete
torsion is possible.  Let us  describe precisely how it is implemented
in conformal field theory, even
though it may be familiar to many readers, as this will facilitate
our discussion of constructing explicit orbifold models with
discrete torsion.
The conformal field theory with target space $M$ is constructed
in terms of maps of $\Sigma\to M$, with $\Sigma$ a Riemann surface.
For the conformal field theory of the orbifold $M/\Gamma$, one must
consider maps of $\Sigma\to M/\Gamma$.
A map to $M/\Gamma$ can be regarded as a map to $M$ which, in looping
around a non-contractible path in $\Gamma$, is ``twisted'' by elements
of $\Gamma$.\foot{Because $\Gamma$ is abelian, the twist involves
a well determined element of $\Gamma$; otherwise one would get only
a conjugacy class.}
For instance, suppose $\Sigma$ is of genus one; for instance,
let $\Sigma$ be the quotient of the $\sigma-\tau$ plane by
$\sigma\to \sigma+m,\,\tau\to\tau+n$, with $m,n\in{\bf Z}$.  A map of
$\Sigma$ to $M/\Gamma$ involves twists by elements of $\Gamma={\bf Z}_k\times
{\bf Z}_k$ in both the $\sigma$ and $\tau$ directions.
If $\zeta$ denotes a generator of ${\bf Z}_k$ -- we will identify
$\zeta$  with the complex number $\exp(2 \pi i/k) $ -- then
the $\sigma$ and $\tau$ twists involve elements of $\Gamma$ that we can
write
\eqn\writet{\eqalign{T_\sigma & = (\zeta^a,\zeta^b)\cr
                     T_\tau & = (\zeta^{a'},\zeta^{b'}).\cr}}
Discrete torsion for $\Gamma$ can now be described explicitly.
Pick an integer $m=0,1,\dots,k-1$.  In the path integral of the orbifold,
weight a sector with given twists $T_\sigma, \, T_\tau$ by an extra
factor
\eqn\ritet{\epsilon(T_\sigma,T_\tau)= \zeta^{m(ab'-ba')}.}
Thus, depending on the choice of $m$, there are $k$ distinct possible
sets of weights in the path integral.  The ``usual'' theory is $m=0$,
and the $k-1$ distinct non-zero choices for $m$ give the theories with
discrete torsion.
The formula \ritet\ has a unique
generalization to genus $g$ that is compatible
with factorization in any channel.  In fact this can be described
once we choose a marking on a Riemann surface involving a canonical
basis of 1-cycles $(a_i,b_i)$.  Relative to this marking the
twisting can be described by thinking of $a,b,a',b'$ above
as corresponding to $g$-dimensional vectors. In that case
the formula \ritet\ for the discrete torsion is still valid
where we think of the products as inner products of vectors.

\bigskip
\noindent{\it Hamiltonian Formulation}

So far we have described the path integral realization of discrete
torsion.  It is also convenient to know how discrete torsion appears
in the Hamiltonian formulation.  Let us think of $\tau$ as ``time''
on the toroidal world-sheet $\Sigma$, and $\sigma$ as ``space.''
The configurations
at $\tau=0$ are classified by the twist $T_\sigma$.  By quantizing
the configurations of given $T_\sigma$, one constructs a Hilbert
space ${\cal H}_{T_\sigma}$; this is called the Hilbert space in
the sector twisted by $T_\sigma$.
In the path integral, in addition to summing over $T_\sigma$, one
also sums over $T_\tau$.  The summation over $T_\tau$ gives a projection
onto the $\Gamma$-invariant part of ${\cal H}_{T_\sigma}$.

{}From this point of view, what is the meaning of the extra factor
\ritet\ in the path integral?  In fact,
the group $\Gamma$ acts on ${\cal H}_{T_\sigma}$,
but this action is not uniquely determined.  There is a ``standard''
action on the Hilbert space that comes from the action of $\Gamma$
on the space of configurations; let us call this standard representation
$T_\tau\to \widehat T_\tau$.   But a new representation can be defined by
\eqn\butby{\widehat T_\tau'=\widehat T_\tau \cdot \epsilon(T_\sigma,T_\tau).}
Indeed, since $\epsilon$ has the property
$\epsilon(x,yz)=\epsilon(x,y)\epsilon(x,z)$, including this factor -- for any
fixed $T_\sigma$ -- still leaves us with a representation of $\Gamma$.
In the path integral with the factor \ritet, the sum over $T_\tau$
corresponds to a projection onto $\Gamma$-invariant states, using
the $\Gamma$ action in \butby.

If ${\cal H}_{T_\sigma}{}^\Gamma$ is the
$\Gamma$-invariant part of ${\cal H}_{T_\sigma}$, then overall the complete
Hilbert space is
\eqn\jurry{{\cal H}=\oplus_{T_\sigma}{\cal H}_{T_\sigma}{}^\Gamma.}

\subsec{An Example}

Now we consider our first real example: a ${\bf Z}_2\times {\bf Z}_2$
orbifold in complex dimension three.

For $i=1\dots 3$, let $z_i$ be a complex variable, $L_i$ a lattice in
the $z_i$ plane, and $E_i={\bf C}/L_i$ the quotient of the $z_i$ plane
by $L_i$; of course, $E_i$ is a Riemann surface of genus one.
Set $T=E_1\times E_2\times E_3$.  Let
$\Gamma$ be the group of symmetries of $T$ consisting of transformations
of the form
\eqn\ofform{z_i\to (-1)^{\epsilon_i}z_i,}
with
\eqn\hohfor{\prod_i(-1)^{\epsilon_i}=1.}
This condition ensures that the holomorphic three-form $\omega=dz_1\wedge
dz_2\wedge dz_3$ is $\Gamma$-invariant.  $\Gamma$ is isomorphic to
${\bf Z}_2\times {\bf Z}_2$, and has three non-trivial elements,
each of which changes the sign of precisely two of the $z_i$.

We wish to consider the Calabi-Yau orbifold $T/\Gamma$.
First we consider some simple facts about the classical geometry.
The operation $z_i\to -z_i$ on the torus $E_i$ has four fixed points.
Using this, it is easy to work out the fixed point sets
of the non-trivial elements of $\Gamma$.  They are all similar, so we
may as well consider the   element $\alpha$ that acts as $-1$
on $z_1$ and $z_2$ and as $+1$ on $z_3$.  Since there are four invariant
values of $z_1$, four invariant values of $z_2$, and the action on $z_3$
is trivial, the fixed point set of $\alpha$ consists of $4\times 4=16$
copies of $E_3$.  Since the two other non-trivial elements of $\Gamma$ act
similarly, the set $W$ on which $\Gamma$ does not act freely is a union of
$3\times 16=48 $ tori.  We will call them fixed tori although each such
torus is fixed only by a ${\bf Z}_2$ subgroup of $\Gamma$.

So far we have classified the points in $T$ that are invariant under
one ${\bf Z}_2$ subgroup of $\Gamma={\bf Z}_2\times {\bf Z}_2$.
The other singular orbits
correspond to points that are left fixed by all of $\Gamma$, in other
words points invariant under $z_i\to -z_i$ for $i=1,2,3$.  The number
of such points is $4\times 4\times 4=64$.  Each of the 64 fixed points
lies at the intersection of three fixed tori; for instance, the fixed
point $z_1=z_2=z_3=0$ is the intersection of the torus $z_1=z_2=0$,
the torus $z_1=z_3=0$, and the torus $z_2=z_3=0$.

\bigskip
\noindent{\it Cohomology}

To understand some essential   properties of this orbifold in string
theory, let us compute the spectrum
of ground states in the Ramond (R) sector\foot{Here
and in the following, we take periodic
boundary conditions for left- and right-moving fermions;
thus what we call the Ramond sector is sometimes called the
Ramond-Ramond or RR sector.}; this gives the analog
of what for a smooth Calabi-Yau manifold would be the cohomology.

First we consider the untwisted sector.  The cohomology of any one
of the $E_i$ is described by the Hodge diamond
\eqn\hodd{\pmatrix{ 1 & 1\cr 1 & 1 \cr}.}
The numbers displayed here are the dimensions of $H^{p,q}(E_i)$,
with $p$ being the vertical axis and $q$ the horizontal axis.
The transformation $z_i\to -z_i$ acts as $+1$ on $H^{0,0}$ and $H^{1,1}$
and as $-1$ on $H^{1,0}$ and $H^{0,1}$.  This follows from the fact
that $H^{0,0}, $ $H^{1,1}$, $H^{1,0}$ and $H^{0,1}$ are generated
respectively by the differential forms $1$, $dz_i\wedge d\bar z_i$,
$dz_i$ , and $d\bar z_i$.

The Ramond ground states coming from the untwisted sector are
simply the $\Gamma$-invariant part of $H^*(E_1\times E_2\times E_3)
=H^*(E_1)\times H^*(E_2)\times H^*(E_3)$.  With $H^*(E_i)$ as described
in the previous paragraph, the $\Gamma$-invariant part of $H^*(E_1\times
E_2\times E_3)$ is readily determined and can be summarized by the Hodge
diamond
\eqn\norfo{\pmatrix{1 & 0 & 0 & 1 \cr
                   0 & 3 & 3 & 0 \cr
                   0 & 3 & 3 & 0 \cr
                   1 & 0 & 0 & 1 \cr}.}
For instance, $H^{1,1}$ is three-dimensional, generated by
$dz_i\wedge d\bar z_i$, for $i=1,2,3$, $H^{2,1}$ is three-dimensional,
generated by such forms as $dz_1\wedge d z_2\wedge d\bar z_3$,
and $H^{3,0}$ is one-dimensional,
generated by the holomorphic three-form $\omega$.

To complete the picture, we must determine the spectrum of
R ground states from the twisted sectors.  There are three twisted
sectors since the abelian group $\Gamma$ has three non-trivial elements.
As we noted above, each non-trivial element acts as $-1$ on precisely
two of the three $z_i$, so we may as well consider a group element
$\alpha$ that acts as $-1$ on $z_1$ and $z_2$ and as $+1$ on $z_3$.
We must find the $\Gamma$-invariant R ground states in the Hilbert
space ${\cal H}_\alpha$ of strings twisted by $\alpha$.

Classically, to have zero energy a string should be a constant configuration,
that is independent of the spatial string coordinate $\sigma$.
In a sector twisted by $\alpha$, the constant must be a fixed point
of $\alpha$.  The element $\alpha$ has four fixed points in acting
on $E_1$, four in acting on $E_2$, and of course acts trivially on $E_3$.
So the fixed point set of $\alpha$ consists of sixteen copies of $E_3$.
Quantization of the space of constant strings gives the space of R ground
states in ${\cal H}_\alpha$; it is a sum of sixteen copies of the cohomology
of $E_3$.  Before writing the corresponding contribution to the Hodge
diamond, that is the dimensions of the $H^{p,q}$, we must remember that
$p$ and $q$ arise in the physics as certain $U(1)$ charges and that
in the twisted sectors the zero point values of $p$ and $q$ receive
certain shifts which are needed to ensure Poincar\'e duality
(or CPT invariance) of the orbifold
theory.\foot{If the complex
coordinates $z_i$
are twisted by $z_i\to e^{i\theta_i}z_i$ with $0\leq \theta_i<2\pi$,
then the zero point shift in $p$ and $q$ is by $\sum_i\theta_i/2\pi$
\orbi.  This has been studied in detail in \ref\zasl{E. Zaslow,
``Topological Orbifold Couplings And Quantum Cohomology Rings,''
Comm. Math. Phys. {\bf 156} (1993) 301.}.
All assertions in this paper about zero point shifts  follow
from this formula.   In the present example, the $\theta_i$
are $\pi,\pi,0$ so the shift is by $1$.}
  The shifts mean that the sixteen copies of $H^{p,q}(E_3)$ contribute
to $H^{p+1,q+1}$ of the orbifold theory, so that the R ground
states in ${\cal H}_\alpha$ can be described by the Hodge diamond
\eqn\jjj{\pmatrix{ 0 & 0 & 0 & 0 \cr
                  0 & 16 & 16 & 0 \cr
                  0 & 16 & 16 & 0 \cr
                  0 & 0  & 0 & 0 \cr}.}

It remains to extract the $\Gamma$-invariant subspace of these twisted
R ground states.  In the absence of discrete torsion, we take
the natural $\Gamma$ action on $H^{p,q}(E_3)$.  Since $\Gamma$ acts
on $E_3$ through $z_3\to -z_3$, in the natural action of $\Gamma$,
the forms
$1$ and $dz_3\wedge d\bar z_3$ are invariant and $dz_3$ and $d\bar z_3$
are not.  So the $H^{1,1}$ and $H^{2,2}$ contributions in \jjj\ survive
as contributions to the ground state spectrum of the orbifold.
Since there are three twisted sectors, obtained by permutations
of the $z_i$ from the one we have analyzed, the total contributions
of the twisted sectors to the dimensions of $H^{1,1}$ and $H^{2,2}$
are $3\times 16 = 48$.  Adding these figures to the untwisted Hodge
diamond of equation \norfo, the Hodge diamond of the orbifold
theory without discrete torsion is
\eqn\wigo{\pmatrix{ 1 & 0 & 0 & 1 \cr
                   0 & 3 & 51 & 0 \cr
                   0 & 51 &  3 & 0 \cr
                   1 &  0 &  0 & 1 \cr}.}

It remains to consider the case with discrete torsion.  Using the explicit
description of discrete torsion for ${\bf Z}_n\times {\bf Z}_n$ in
\ritet, it can be seen that for the $\Gamma$ element $\alpha$ that
we have considered, $\epsilon(\alpha,\beta)=-1$ precisely if
$\beta\not=1,\alpha$.
With the particular ${\bf Z}_2\times {\bf Z}_2$ action that we have
taken on $T=E_1\times E_2\times E_3$, $\beta$ acts on $z_3$ as
$z_3\to -z_3$ precisely if $\beta\not= 1,\alpha$.  Putting these
facts together, the effect of the discrete torsion in the sector
twisted by $\alpha$ is simply to include
an extra minus sign in the transformation of the states under $z_3\to -z_3$.
With the new transformation law, $dz_3$ and $d\bar z_3$ are even
and $1,dz_3\wedge d\bar z_3$ are odd.
So with discrete torsion, the part of \jjj\ that contributes
to the cohomology of the orbifold is $p=2,q=1$ and vice-versa.
Including the three twisted sectors, the cohomology of the orbifold
is therefore described by the Hodge diamond
\eqn\uwigo{\pmatrix{ 1 & 0 & 0 & 1 \cr
                   0 & 51 & 3 & 0 \cr
                   0 & 3 &  51 & 0 \cr
                   1 &  0 &  0 & 1 \cr}.}

Notice that \wigo\ and \uwigo\ are mirrors of each other, that
is related by $p\leftrightarrow 3-p$, $q\leftrightarrow q$; this is
the operation that reverses the sign of the left-moving $U(1)$ charge
without affecting the right-moving one.  This hints that the theories
with and without discrete torsion are mirror; this is true, as we will
explain in section 3.

With or without discrete torsion, the twisted sectors added 48 states
to $H^{1,1}$ or $H^{2,1}$.  Each such state arose
as the contribution of one of the 48 fixed tori in $T$.  Each of these
tori becomes a ${\bf Z}_2$ orbifold singularity in $T/\Gamma$.  These
singularities are of complex codimension two
and look locally like the singularity
\eqn\riddo{y^2=uv.}  This singularity can either be deformed away,
by adding a parameter to the equation to give, say,
\eqn\iddo{y^2=uv+\epsilon,}
or it can be blown up.  In terms of the three-dimensional orbifold theory,
the deformation involves a mode in $H^{2,1}$ supported along the fixed
torus, and the blowup involves a mode in $H^{1,1}$ that is  supported there.

{}From our computation of the spectrum, it is clear what is happening.
Without discrete torsion, the twisted sector modes are in $H^{1,1}$ and
the fixed torus
singularity is blown up; with discrete torsion, the twisted sector
modes are in $H^{2,1}$ and the fixed torus singularity is deformed.

Blowing up or deformation of the fixed tori removes the singularities
of the orbifold $T/\Gamma$
in complex codimension two.  But what happens to the 64 fixed points
-- is one left with singularities in complex codimension three?
In the case of the blow-up -- that is, without discrete torsion
-- the answer is that blowing up the codimension two singularities
automatically also eliminates the singularities in codimension three.
This is related to the fact that for abelian groups, the conformal
field theory of orbifolds agrees with classical geometry \aspin,\foot{The
analysis of \aspin\ does not quite apply to the present ${\bf Z}_2\times {\bf
Z}_2$ example, because of the codimension two singularities, but
the extension to the ${\bf Z}_2\times {\bf Z}_2$
example has been described by the author
of \aspin.}  as a result of which in this paper we
will have little to say about the theory without discrete
torsion.

With discrete torsion, one encounters deformation rather than
blowup, and  the answer is
quite different, as we will see.

\subsec{Deformation Of The Orbifold}

We wish  to compare the orbifold
conformal field theory with discrete torsion
to the classical geometry obtained
by  deforming the singularities of the orbifold $T=(E_1\times E_2\times E_3)
/\Gamma$.

An alternative description of
the genus one surfaces $E_i$ is convenient.
A Riemann surface $E$ of genus one can be
described as a double cover of ${\bf CP}^1$ branched over four points;
it can be described, therefore, by an equation $y^2=F(u,v)$ where
$u,v$ are homogeneous coordinates of ${\bf CP}^1$, $y$ is homogeneous
of degree 2,  and $F$ is a homogeneous
quartic polynomial.  ${\bf Z}_2$  acts on $E$ by $y\to -y$, with $u,v$
invariant.  The fixed points of ${\bf Z}_2$ are therefore the four
homogeneous solutions of $F(u,v)=0$.

So $T=E_1\times E_2\times E_3$ can be described by
variables $u_i,v_i,y_i$, $i=1\dots 3$, with
equations
\eqn\wieq{y_i{}^2=F_i(u_i,v_i).}
To give an algebraic description of the quotient $T/\Gamma$,
we simply identify the $\Gamma$-invariant sub-ring of the ring of
polynomial functions in the $u_i,v_i$, and $y_i$.  Noting that $\Gamma$
acts by pairwise sign changes of the $y_i$, the invariants are the $u_i$,
the $v_i$, and $y=y_1y_2y_3$ ($y$ is homogeneous of degree two with respect
to scalings of any pair $u_i,v_i$), subject to the one equation
\eqn\tieq{y^2=\prod_{i=1}^3 F_i(u_i,v_i).}

Now -- leaving physics aside -- it is clear how to deform the complex
structure of $T/\Gamma$ to get a smooth Calabi-Yau manifold.  One simply
deforms the function $\prod_iF_i(u_i,v_i)$ to a generic polynomial
$F(u_1,v_1;u_2,v_2;u_3,v_3)$ which is homogeneous of degree four in each pair
of variables $u_i,v_i$.  The equation
\eqn\yieq{y^2=F(u_i,v_i)}
describes a double cover of ${\bf CP}^1\times {\bf CP}^1\times {\bf CP}^1$
which -- for generic $F$ -- is a smooth Calabi-Yau manifold.

Let us count the number of parameters in this family of Calabi-Yau manifolds.
The space of quartic polynomials in two variables $u,v$ is of dimension 5.
The space of polynomials $F$ of degree four in each pair is therefore
of dimension $5\times 5\times 5=125.$  From this number we should remove
$3\times 3=9$ corresponding to the action of $SL(2,{\bf C})$ on each pair
$u_i,v_i$, and 1 for overall scaling of $F$ (which can be absorbed in scaling
of $y$).  So the total number of polynomial deformations is $125-9-1=115$.
Though more complicated phenomena occur for other Calabi-Yau manifolds
(see \ref\hubsch{T. Hubsch, {\it Calabi-Yau Manifolds:
A Bestiary For Physicists} (World Scientific, 1992).},
Chapter A, for an introduction),
in the present example, it can be readily shown that the polynomial
deformations faithfully represent the possible deformations of the complex
structure.

\subsec{The Meaning Of The Discrepancy}

{}From these
calculations, we get a discrepancy between the orbifold with discrete
torsion and the classical geometry.
The orbifold with discrete torsion is part of
a family of conformal field theories that depends on 51 ``complex structure''
moduli, while the smooth Calabi-Yau manifold that is obtained by deformation
of complex structure of $T/\Gamma$ has 115 complex structure moduli.
The difference is $115-51=64$.\foot{On the other hand, it can be seen,
for instance by computing the Euler characteristic, that the smooth
Calabi-Yau manifold given by a generic equation \tieq\ has $b^{1,1}=3$,
in agreement with the conformal field theory of the orbifold with discrete
torsion.}

But
64 is a number that we have already seen: it is the number of ${\bf Z}_2\times
{\bf Z}_2$ fixed points.  As we have already discussed, these
are the points that may remain as codimension three singularities
after the codimension two singularities are deformed away.

Let us first explain the interpretation that we wish to propose
for the discrepancy.
A generic equation \yieq\ depends on 115 parameters and describes
a smooth Calabi-Yau manifold.  Suppose that one does not wish to get
a smooth Calabi-Yau manifold but one with certain singularities.
Then restrictions must be placed on the parameters, so the most
general Calabi-Yau manifold in this family with specified singularities
will depend on fewer than 115 parameters.  The precise number depends
on the type of singularities one prescribes.

\nref\vw{C. Vafa and N. Warner, ``Catastrophes And The Classification
Of Superconformal Theories,'' Phys. Lett. {\bf B218} (1989) 51.}
\nref\martinec{E. Martinec, ``Algebraic Geometry And Effective
Lagrangians,'' Phys. Lett. {\bf B217} (1989) 431.}
Consider a singularity of a three-dimensional complex manifold
described by a general equation
$F(x_1,x_2,x_3,x_4)=0$, with an isolated singularity at the origin, where
 $F=\partial_iF=0$.
Suppose one considers an arbitrary small
perturbation to a nearby equation
\eqn\nerbeq{F(x_1,x_2,x_3,x_4)=\epsilon(x_1,x_2,x_3,x_4).}
Terms in $\epsilon$ that can be expressed as linear combinations
(with holomorphic coefficients) of the partial derivatives $\partial_iF$
can be transformed away by shifting the $x_i$.  (For instance,
$\epsilon=\epsilon_0\partial_1F$ is removed to first order
by $x_1\to x_1+\epsilon_0$.)  So the space of relevant operators
is the space of polynomials in the $x_i$ modulo the ideal generated by the
derivatives $\partial_iF$.  (This is familiar to string theorists
in the Landau-Ginzburg theory of singularities \refs{\vw,\martinec}.)
The identity operator is always relevant. There are additional relevant
operators unless the $x_i$ are all in the ideal, which happens precisely
if $F$ is equivalent locally to the conifold
\eqn\cconi{F=x_1{}^2+x_2{}^2+x_3{}^2+x_4{}^2.}
Thus, the conifold is the unique isolated singularity with precisely
one relevant operator.  If one wishes to have precisely two relevant
operators, then the relevant operators must be 1 and a linear function
\foot{By a linear function we mean really a function
with only a first order zero; it becomes a linear function if coordinates
are chosen correctly.  If there are two relevant operators, one must be linear
since if the $x_i$ are all in the ideal, so are all higher order polynomials
and the identity is the only relevant operator.}
of the $x_i$, say $x_1$; the ideal must then
contain $x_2,x_3,x_4$, and $x_1{}^2$ (or there would be more than two relevant
operators).  One must then have up to a choice of coordinates
\eqn\impo{F=x_1{}^3+x_2{}^2+x_3{}^2+x_4{}^2.}

In a suitably generic
family of Calabi-Yau manifolds, a singularity with $k$ relevant
operators will appear in complex codimension $k$, since to obtain that
singularity one must adjust $k$ relevant parameters.
For instance, a  conifold
singularity will (generically) arise in complex codimension one.
Thus,  if one wishes to deform \tieq\ to an equation describing an
(otherwise generic) Calabi-Yau manifold with a conifold point, the number
of parameters will be 114 instead of 115.
If one wants $n$ disjoint conifold singularities, the number of parameters
is $115-n$.  For $n$ singularities each with $k$ relevant operators,
the number of parameters will be $115-kn$.

In our problem, we start with 48 fixed tori
and 48 twisted sector marginal operators that represent deformations
of those codimension two singularities.  This leaves unclear whether there
will be codimension three singularities at the 64 fixed points.
  If so, the number of complex structure parameters of the Calabi-Yau
will be $115-64n$, where $n$ is the number of relevant operators
of the singularity.  (With appropriate complex structures on the $E_i$,
there are symmetries permuting the fixed points, so $n$ is the same for each.)
The actual number of complex structure deformations
of the conformal field theory with discrete
torsion is $51=115-64$, so this will fit if $n=1$.
But $n=1$ corresponds precisely to the conifold.  Thus, we get a candidate
for the geometrical interpretation of the conformal field theory of the
orbifold with discrete torsion: it corresponds to a Calabi-Yau with
64 conifold singularities.

This is a conifold that cannot be eliminated by any marginal operator
of the conformal field theory, since the 64 requisite operators are missing.
It is as if
discrete torsion supported at the conifold singularity prevents
it from being deformed.  We will
discuss the physical interpretation more fully at the end of this section.

\bigskip
\noindent{\it Local Behavior}

To support our explanation of the apparent discrepancy between
conformal field theory and geometry,
we will examine some of the above-mentioned
ingredients more carefully.  First we look at
the behavior near one of the fixed points.

In terms of the description of the $E_i$ by  equations
$y_i{}^2=F_i(u_i,v_i)$, we
 can take the fixed point to be at $u_1=u_2=u_3=0$.  By scaling, one
can set $v_1=v_2=v_3=1$.  The fixed points are zeroes of the $F_i$,
so we can assume that near $u_i=0$, $F_i\sim u_i$.  Thus
equation \tieq\ takes the form
\eqn\hieq{y^2=u_1u_2u_3}
near the fixed point.
Note that in \hieq, there are  codimension two singularities on
three curves $C_i$ -- one with
$u_1=u_2=0$, another with $u_1=u_3=0$, and another with $u_2=u_3=0$.
The fixed point at $u_1=u_2=u_3=0$ is the intersection of these three
lines of codimension two singularities.  Of course, this is expected
 since more globally the fixed
points are intersections of three fixed tori.

Now the first point is that it is possible to add a perturbation to \hieq\
that eliminates the codimension two singularities but leaves a conifold
singularity at the origin.  This could be simply
\eqn\hulfo{y^2=u_1u_2u_3+\epsilon(u_1{}^2+u_2{}^2+u_3{}^2).}
A general first order perturbation
\eqn\yulfo{y^2=u_1u_2u_3+\epsilon(u_1,u_2,u_3)}
removes the codimension two singularities
if $\epsilon$ is generically non-zero on the $C_i$.  It in addition
removes the singularity at the origin -- where the $C_i$ intersect --
if $\epsilon(0,0,0)\not=0$.

\bigskip
\noindent{\it Global Story}

Now let us examine precisely which 64 modes are missing.
We start with the unperturbed equation
\eqn\pieq{y^2=F_1(u_1,v_1)F_2(u_2,v_2)F_3(u_3,v_3)}
with the $F_i$ being homogeneous quartic polynomials.

Each $F_i$ takes values in a five dimensional space $V_i$ of homogeneous
quartics.  $F_i$ itself generates a one dimensional subspace $V_{i,0}$ of
$V_i$; let $W_i$ be a four dimensional complement to $V_{i,0}$.  We
will take the symbol $\delta F_i$ to refer to a variation of $F_i$
that is constrained to lie in $W_i$.

We perturb
\pieq\ to an equation of the form
\eqn\ppieq{y^2=F_1F_2F_3+ \epsilon\delta F.}
The possible $\delta F$'s can be classified as follows.

There is one variation in which the change in $F_1F_2F_3$ is a multiple
of itself.  This is irrelevant since it can be absorbed in rescaling $y$.
Our notation does not even permit us to write this mode conveniently
since we require $\delta F_i$ to take values in the complementary space $W_i$.

There are twelve deformations of the form
$\delta F=\delta F_1\, F_2F_3+F_1\delta F_2\, F_3+F_1F_2\delta F_3$
in which only one of the $F_i$ is deformed.  However $3\times 3=9$
of them can be removed by $SL(2,{\bf C})$ transformations on $(u_i,v_i)$.
Altogether, then, there are $12-9=3$ non-trivial  modes of this kind.
These deformations give equations that in first order in $\epsilon$
are equivalent to  $y^2=(F_1+\epsilon
\delta F_1)(F_2+\epsilon\delta
F_2)(F_3+\epsilon\delta F_3)$.  This is of the general form of \tieq\ with
different $F_i$, so it describes an orbifold $T/\Gamma$ with a different
complex structure on $T$.  Comparing to conformal field theory,
these are the three modes from the untwisted sector that preserve
the orbifold structure.

There are 48 modes in which two of the $F_i$ are varied.
These are modes of the form
\eqn\gieq{\delta F=\delta F_1\delta F_2\, F_3+F_1\delta F_2\delta F_3
+\delta F_1\, F_2\delta F_3.}
These modes have the property that they vanish on the 64 fixed points
(which are characterized by $F_1=F_2=F_3=0$) but they do not generically
vanish on the fixed tori (on which only two of the $F_i$ vanish).
Therefore, the modes of this form remove the codimension two singularities
but in first order leave codimension three singularities at the
fixed points.

Finally, there are 64 modes in which all three $F_i$ are varied,
\eqn\ggieq{\delta F=\delta F_1\delta F_2\delta F_3.}
These modes have no particular zeroes and would remove all the singularities.
We claim, however, that these are the modes that are missing in the conformal
field theory.

This is strongly suggested by the structure of the computation
of the twisted sector modes in the conformal field theory.
Each twisted sector mode comes from a sector twisted by a group element similar
to the element $\alpha$
that acts non-trivially on the two elliptic curves
$E_1$ and $E_2$ and trivially on $E_3$.  The modes in the
$\alpha$-twisted sector should deform the singularity of the $\alpha$-fixed
tori, that is the singularities with $F_1=F_2=0$.  They should not deform
the singularities from fixed tori of other group elements.  The modes
that do this must vanish when $F_1=F_3=0$ or $F_2=F_3=0$
(so as not to disturb the fixed tori of other group elements) but
not when $F_1=F_2=0$.  These modes must therefore be of the
form  $\delta F_1\delta F_2F_3$.
Similarly,  the other twisted sectors give the other modes in \gieq.
But nothing in the conformal field theory gives the  64 modes in \ggieq.

\bigskip
\noindent{\it Support Of The Torsion}

Before discussing the support of the torsion, we need to recall
a generality about discrete torsion.  In discrete torsion, one starts
with an element $\gamma$ of $H^2(\Gamma,U(1))$, and (as long as one
keeps away from singularities) this is then mapped to an element
$\widehat\gamma$ of $H^2(M_0,U(1))$, with $M_0$ the smooth part of
the orbifold $T/\Gamma$; physically, the world-sheet theory then
has a $B$-field in the cohomology class of $\widehat \gamma$.
  If $\widehat \gamma$ is in the connected
component of $H^2(M_0,U(1))$, this $B$ field is described as a world-sheet
theta angle; if not, it is called discrete torsion in space-time.
Both possibilities can arise \ref\morasp{P. Aspinwall and D. Morrison,
``Chiral Rings Do Not Suffice: $N=(2,2)$ Theories With Nonzero
Fundamental Group,'' DUK-TH-94-70, IASNSN-HEP-94/37.}.

Returning to our problem,
let $p$ be a fixed point in the orbifold $T/\Gamma$.  The underlying discrete
torsion is non-trivial in an arbitrarily small neighborhood of $p$ since
it appears in the weight of the path integral for string world-sheets
that sit arbitrarily close to (or even at) the singularity.

After the complex deformation is made, in the presence of the underlying
discrete torsion,
a neighborhood of $p$ looks like the complex singularity
\eqn\pilop{x_1{}^2+x_2{}^2+x_3{}^2+x_4{}^2=0}
with $p$ being the point $x_i=0$.  If one deletes the conifold point $p$
from this space, one gets a smooth manifold $W$
with the homology of ${\bf S}^2
\times {\bf S}^3$.\foot{If one treats the variables in \pilop\ as homogeneous
variables, the equation describes a smooth quadric in ${\bf CP}^3$, isomorphic
to ${\bf CP}^1\times {\bf CP}^1$.  In projectivizing the variables, one
divides by a ${\bf C}^*$ action, so the conifold with the origin deleted
is a ${\bf C}^*$ bundle over ${\bf CP}^1\times {\bf CP}^1$.  The
cohomology can be computed using this fibration.}
In particular, $H^2(W,{\bf Z})\cong {\bf Z}$ and $H^2(W,U(1))\cong U(1)$.
As this is connected, the underlying discrete torsion of the orbifold
could not produce discrete torsion of $W$, but it might produce a theta
angle.  To test this possibility, we need to know whether the discrete
torsion produces a phase for a string world-sheet that wraps around
a generator ${\bf S}$ of $H_2(W,{\bf Z})\cong {\bf Z}$.
One can choose ${\bf S}$ to be a two-sphere obtained in
resolving the fixed tori of the orbifold; in that operation
a point with $z_1\not=0$, $z_2=z_3=0$ is replaced by a two-sphere ${\bf S}$.
But for a world-sheet with $z_1$ almost constant and non-zero,
the discrete torsion (which only contributes in sectors in which all
three $z_i$ are twisted non-trivially) does not give any phase.
Therefore, the underlying discrete torsion of the orbifold does
not contribute anything if restricted to $W$.

To summarize all our statements, the underlying discrete torsion
produces an effect which is non-zero
in an arbitrarily small neighborhood of $p$, but zero if $p$ is deleted;
it is roughly as if the discrete torsion has delta function support at $p$.
One can roughly model the situation by supposing that, with the proper
definition, the conifold $M$ has a torsion element in $H^2(M,U(1))$
which would disappear if the conifold singularity is deformed or
resolved; then discrete torsion supported at the conifold point
would explain the inability to deform the conifold.  However, we do
not know the proper definition of $H^2(M,U(1))$ to justify this interpretation.

In any event, what is going on at the conifold singularity can not
necessarily be properly interpreted as a discrete torsion with delta
function support.  It is not at all clear that the conifold theory
that we have found, which does make sense, differs just by discrete phases
from the more traditional (singular) $A$ and $B$ models of the conifold.
Our argument started with an orbifold with discrete torsion, which differed
from the ordinary orbifold only by such phases, but by the time we
deform to the conifold, there is no way to compare the model
to another model that is ``identical except for phases.''

As another interpretation, perhaps in an infinitesimal neighborhood
of the singularity an $H$-field is turned on (recall $H=dB$ where
$B$ is the two form).   This may be natural from the following viewpoint.
If there were no fixed points, inclusion of discrete torsion
is equivalent to turning on a $B$-field.  Just as the orbifolds
have curvature singularities concentrated at the fixed points, it may be
that orbifolds with discrete torsion
have the torsion field $H$ concentrated
at the fixed points.   Moreover this may also explain why the singularity
cannot be deformed.  It can be shown that with an $H$ field turned on
in a smooth way the $N=2$ superconformal symmetry is broken \ref\rowi{
R. Rohm and E. Witten, ``The Antisymmetric Tensor Field In String Theory,''
Ann.Phys. {\bf 170} (1986) 454.}.  It may be the case
that with
delta function support
the $N=2$ superconformal
symmetry may be restored.
In fact this may be the reason that in the context of $N=2$
superconformal theory with torsion there is no marginal deformation that gets
rid of the singularity.

A similar
phenomenon occurs in the context of bosonic string orbifolds.
In the case of bosonic strings,
conformal invariance for smooth manifolds
favors that the manifold should be flat.  This however is in contrast with
explicit toroidal orbifold conformal theories
 which are flat everywhere except for
delta function curvature singularities.  One would thus expect
that in the case of bosonic string there are no marginal operators
that get rid of singularities of orbifolds;  this is indeed generally the case.
Thus bosonic orbifolds provide examples of isolated singularities
that cannot be deformed--very much as discrete torsion on the conifold
behaves in the superstring case.

To summarize this, all we really know is that the underlying
discrete torsion produces an effect -- an $H$ field or something else --
that is supported at the singularity
and whose presence makes the singularity inescapable.
It would be very interesting
to find a more
 explicit description of this  quantum field theory and learn what
is going on.

\newsec{Mirror Symmetry and ${\bf Z}_2\times {\bf Z}_2$ Orbifold with Torsion}

In this section we will show that the ${\bf Z}_2\times {\bf Z}_2$ orbifold
described in the previous section provides a simple realization of
mirror symmetry: the orbifold with discrete torsion is mirror to the
same orbifold without discrete torsion.
We have already found a hint of this in observing that the Hodge
diamonds of these models are mirror of one another.  Here we will
show that they indeed are identical superconformal theories.
This implies, in particular, that the complex structure
deformation of the ${\bf Z}_2\times {\bf Z}_2$ orbifold with
discrete torsion described in the last section is mirror
to the Kahler deformation of the same orbifold theory without torsion.
This also implies that by studying the periods of the deformation discussed
there one should be able to deduce the quantum
cohomology ring for the blown-up orbifold.  This computation
should be interesting
to do as it would provide a further check on the
geometrical interpretation of the orbifold with discrete torsion
advanced in the last section.

In order to show that the two orbifold theories with and without
discrete torsion are equivalent one should show an identification
between the operators of the two theories under which the correlation
functions and partition functions coincide.  It is generally considered
sufficient to
show that the spectra are  identical in the two cases and
that in addition they have the same three point correlators
on the sphere, or alternatively,
that the partition functions at all genera are identical.
In the case at hand, all these things are easy to prove.

\subsec{Mirror Symmetry and Phase of Path-Integral}

Mirror symmetry is an isomorphism between two $(2,2)$ conformal field
theories under which the sign of the left-moving $U(1)$ charge, but
not the right-moving one, is reversed.  Roughly, this means that the
complex structure is reversed for left-movers but not for right-movers.
As is well known, in the  case of a three-fold
such a symmetry
flips the sign of ${\rm Tr}(-1)^{\rm F}$ for the supersymmetric
ground states.

A basic result in this area is that the mirror of a complex torus is
another complex torus.  Let us recall how this comes about.  Begin
by considering a free scalar field $Y$, compactified on a circle of
radius $R$.  It is well known that this theory is equivalent to a similar
theory with radius $1/R$; under this transformation there is a non-trivial
identification of operators:
\eqn\nontri{\eqalign{\partial Y & \to \partial Y\cr
                     \bar\partial Y & \to -\bar\partial Y.\cr}}

Now add a second free periodic boson $X$, of radius $R'$.  We suppose
that the metric is just $ds^2=dX^2+dY^2$ so that a local complex
coordinate is just $Z=X+iY$.  Consider the transformation $R\to 1/R$ for
the $Y$ variable, without doing anything to $X$.  (The $B$ field
must vanish to make this a symmetry.) Obviously, as
the operators $\partial X$ and $\bar\partial X$ are invariant, one has
\eqn\bontri{\eqalign{\partial Z & \to \partial Z\cr
                     \bar\partial Z &\to \bar\partial\, \bar Z.\cr}}
Thus, the complex structure is reversed for left-movers but not for
right-movers; this is a mirror symmetry.  Note in particular that
the volume ($R R'$) and the shape ($R'/R$) get exchanged
under $R\to 1/R$.
Thus a two dimensional
rectangular torus
with zero $B$ field and radii $R',R$ is mirror to a similar model
with radii $R',1/R$.  By following the possible deformations on the two
sides, one learns that any two-torus is mirror to another two-torus.

There is no problem in generalizing this to higher dimensions.
One simply considers several periodic variables $X_i$, several $Y_i$,
and $Z_i= X_i+iY_i$.  A transformation
$R\to 1/R$ on each of the $Y_i$ brings about
\eqn\contri{\eqalign{\partial Y_i&\to \partial Y_i \cr
\bar\partial Y_i & \to -\bar\partial Y_i,\cr}}
and so
\eqn\bontri{\eqalign{
\partial Z_i & \to \partial Z_i\cr
                     \bar\partial Z_i&\to \bar\partial \,\,\bar Z_i.\cr}}
So this transformation is a mirror symmetry.

Now, let us consider precisely how this mirror symmetry acts on
the fermions.  Each $X_i$ and each $Y_i$ is related by world-sheet
supersymmetry to a right-moving fermion and to a left-moving fermion.
Since the right-moving parts of $X_i$ and $Y_i$ are invariant under
the mirror symmetry, world-sheet supersymmetry implies that
the right-moving fermions are also invariant.
The same is true for the left-moving
partners $\psi_i$
of $X_i$.  However, as the left-moving part of $Y_i$ has
its sign reversed under $R\to 1/R$, the left-moving partners $\eta_i$ of $Y_i$
change sign in this operation.  Thus, if we are in complex dimension
three, the mirror symmetry reverses the sign of precisely three
left-moving fermions.

In fact, this lets us check that this operation is a mirror symmetry.
The operator $(-1)^{F_L}$ in the Ramond sector (the left-moving part
of $(-1)^F$) contains a zero mode part that is the product of the zero
modes of $\psi_i$ and $\eta_i$.  As there are an odd number of $\eta$'s,
the reversal in sign of the $\eta_i$ gives the expected sign change
of $(-1)^{F_L}$ under mirror symmetry.

\bigskip
\noindent{\it Path Integral Formulation}

Let us discuss what this looks like from a
path integral point of view.  On world-sheet fermions, we may impose
antiperiodic $(A)$ or periodic $(P)$ boundary
conditions in circling around a
string.  This leads to four boundary conditions or spin structures
for genus 1 depending on the boundary conditions in the $\sigma$ and $\tau$
directions; we may call these $(P,A)$, $(A,A)$, $(A,P)$, and $(P,P)$.
Of these, $(P,P)$ is modular invariant, and the other three are
permuted by modular transformations.  In genus $g$ there are $2^{2g}$
spin structures.

A spin structure is said to be even or odd depending on whether
the number of negative (or positive) chirality
fermion zero modes is even or odd.  For instance, in genus
one, $(P,P)$ is odd (there is one zero mode, the constant), and the
others are even (there are no zero modes).

The importance for us of counting the number of zero modes is that
this determines the behavior of the world-sheet path integral
measure under mirror symmetry.  We have seen that mirror symmetry
acts by $\eta_i\to -\eta_i$ (with other fermions invariant), so the measure
is even or odd depending on whether the number of components of $\eta_i$
is even or odd.  The non-zero modes are naturally paired, so the measure
is even or odd depending on whether the number of zero modes of the $\eta_i$
is even or odd.  If therefore $\alpha$ denotes the spin structure of
left-moving fermions and
$\sigma_\alpha$ is 0 or 1 for $\alpha$ an even or odd spin structure,
then for target space a torus the genus $g$
measure $\mu_{g,\alpha}$
with spin structure $\alpha$
transforms under mirror symmetry as
\eqn\pillop{\mu_{g,\alpha}\to (-1)^{\sigma_\alpha}\mu_{g,\alpha}.}

\bigskip
\noindent{\it The Orbifold}

Now we consider a situation with three $Z_i=X_i+iY_i$, and we introduce
the group $\Gamma={\bf Z}_2\times {\bf Z}_2$, acting, as in section 2,
by pairwise sign changes of the $Z_i$.

Since the group $\Gamma$ commutes with the transformation \contri\ of
the $Y_i$ under $R\to 1/R$, the $R\to 1/R$ operation can be done
for the orbifold $T/\Gamma$, not just for the original torus $T$.
However, just as for the original torus, the $R\to 1/R$ transformation
induces a sign change $\eta\to   -\eta$, and we must determine what
this sign change
does to the path integral measure.  We will do this explicitly in genus
one before discussing the generalization.

To do the genus one
path integral of the orbifold, we have to consider the path
integral with toroidal target and various
twisted boundary conditions in the $\sigma$ and $\tau$ directions.
Let $g$ and $h$ be the elements of $\Gamma$ that act on $(Z_1,Z_2,Z_3)$
as $(1,-1,-1)$ and $(-1,1,-1)$, respectively.
A general twist would involve a pair of group elements $(x,y)=(g^rh^s,g^th^u)$,
and as stated in equation \ritet, the discrete torsion for this
pair corresponds to a sign factor
\eqn\pimmo{\epsilon(g^rh^s,g^th^u)=(-1)^{ru-st}.}
We want to show that for fermions with a given spin structure $\alpha$
and a given set of ${\bf Z}_2\times {\bf Z}_2$ twists $(x,y)$, the
path integral measure transforms under $R\to 1/R$ by
\eqn\immo{\mu_\alpha(g,h)\to (-1)^{\sigma_\alpha}\epsilon(x,y)\mu_\alpha(g,h).
}
If true, this asserts that the orbifold theory without (or with)
discrete torsion transforms under $R\to 1/R$ into the mirror theory
with (or without) discrete torsion.  The factor $(-1)^{\sigma_\alpha}$
makes the transformation a mirror transformation, and the second
factor is the discrete torsion.

Up to modular transformations
and permutations of the $Z_i$, it is enough to check \immo\ for
$(x,y)=(1,1)$, $(g,1)$, and $(g,h)$.  We already know that the result
is true for $(1,1)$.  Let us consider $(g,h)$.

Suppose that $\sigma_\alpha=1$, that is, suppose that the fermions
are in the odd or periodic spin structure $(P,P)$.  The
$(g,h)$ twist reverses the boundary conditions for the $\eta_i$
(supersymmetric partners of the $Y_i$) in the $\sigma $ or $\tau$ directions
and effectively shifts the $\eta_i$ into the even spin structures
$(P,A)$, $(A,P)$, and $(A,A)$.
So the measure with $\sigma_\alpha=1$ and twists $(g,h)$ is even;
this agrees with \immo.

Now keep the twists $(g,h)$ but take
$\sigma_\alpha=0$.  It suffices to consider the spin structure $(A,A)$
as the other two even spin structures are related to this by a modular
transformation (which preserves $(g,h)$ up to a permutation of the $Z_i$).

In this case, the effect of the $(g,h)$ twist is to shift
$\eta_3$ into the odd spin structure while leaving the others in
even spin structures.  So, as one fermion is in an odd spin structure,
the path integral measure is odd, as predicted by the above
formula, for even spin structure and
twists $(x,y)=(g,h)$.

To complete the verification of \immo, it remains to consider the
case of twists $(g,1)$.  In this case, $\epsilon=1$ and we must show
that the measure transforms as $(-1)^{\sigma_\alpha}$.  For instance,
for the $(P,P)$ spin structure,
$\sigma_\alpha=1$, the $(g,1)$ twist leaves $\eta_1$ with
effective $(P,P)$ couplings and shifts the others to $(A,P)$; hence
there is one fermion zero mode and the measure is odd under $\eta\to -\eta$.
For spin structure $(A,P)$, $\sigma_\alpha=0$, the twist by $(g,1)$ puts
an even number of fermions ($\eta_2$ and $\eta_3$) in the odd spin structure,
so the measure is even; for
$(P,A)$ and $(A,A)$, the $(g,1)$ twist leaves all fermions in even spin
structures, so the measure is again even.  This completes the verification
of \immo.

So we have verified that for the orbifold, $R\to 1/R$ on the $Y_i$ has
the following characteristics:
(i) it is  a mirror operation, because of the factor of $(-1)^{\sigma_\alpha}$;
(ii) it
adds (or removes) discrete torsion, because of the factor of
$\epsilon$.
Having understood how the measure transforms under $R\to 1/R$, there
is almost nothing to add to discuss correlation functions.   If operators
${\cal O}_i$ transform into operators ${\cal O}'_i$ under $R\to 1/R$,
then the correlation functions of the ${\cal O}_i$ without discrete torsion
are equal to the correlation functions of the ${\cal O}'_i$ with discrete
torsion.  This is true in genus one from the above analysis of the measure,
and it is true more trivially in genus zero where there is only
an even spin structure.  To verify that the assertion is true in higher
genus, we will presently analyze the path integral measure in arbitrary genus.

\bigskip
\noindent{\it Generalization To Genus $g$}
\def\ba{{\bf A}}
\def\bb{{\bf B}}
\def\bc{{\bf C}}
\def\bd{{\bf D}}

To determine how the measure transforms in genus $g$,
we will need some further facts about spin structures in higher genus
(see \ref\amv{L. Alvarez-Gaume, G. Moore and C. Vafa, ``Theta Functions,
Modular Invariance, and Strings,''
Comm. Math. Phys. {\bf 106} (1986) 40.}\
for a review).  Consider a genus $g$ Riemann surface $\Sigma^g$.
Choose a canonical basis of the 1-cycles labeled by $(a_i,b_i)$ where
$i=1,...,g$.  Such a ``marking''
determines a canonical spin structure.
Relative to this marking, any other spin structure
is determined by twisting the fermions by signs $\pm 1$ around the various
$a_i$ and $b_j$ cycles.

Spin structures can therefore be classified by ${\bf Z}_2$-valued quantities
$\theta_i,\phi_j$, $i,j=1\dots g$, as follows: in the
spin structure $\alpha=(\theta_i,\phi_j)$, fermions are twisted
by an extra minus sign (relative to the canonical
spin structure determined by the marking)

in going around any cycle $a_i$ or $b_j$ such that $\theta_i$ or $\phi_j$
is 1.
It is sometimes convenient to think of $\theta_i ,\phi_i$ as $g$-dimensional
vectors which we label by ${\Theta}, {\Phi}$.
A classic formula says that the parity of the spin structure $\alpha$ is
\eqn\sis{\sigma_\alpha =\sigma (\Theta ,\Phi )=
{\Theta }\cdot {\Phi} \qquad {\rm mod} \ 2}
That is, the number of fermion zero modes in the spin structure $\alpha$
is equal mod two to $\Theta\cdot \Phi=\sum_i\theta_i\phi_i$.

Now we consider the path integral of the orbifold theory for a given
spin structure $\alpha=(\Theta ,\Phi )$, with twists
around the $a_i$ cycle given
by $g^{A_i}h^{C_i}$ and twists around the $b_i$ cycle
given by $g^{B_i}h^{D_i}$.  The $A_i,B_i,C_i,D_i$ are integers
defined mod $2$; we combine them in
vectors ${\bf A},{\bf B},{\bf C},{\bf D}$.
The discrete torsion in this case is given according to equation
\ritet\ by
\eqn\ditor{\epsilon =(-1)^{{\bf A\cdot D}-{\bf B\cdot C}}}
As above, we want to show that under $\eta\to -\eta$, the path integral
measure transforms by
\eqn\mutu{\mu\to \mu \cdot (-1)^{\sigma(\Theta,\Phi)}\epsilon.}

Using the definition of $g$ and $h$,
the effect of the twist is to shift the effective spin structures for
the fermions in the $Z_1, Z_2,Z_3$ directions to
$$Z_1:\qquad (\Theta +{\bf C},\Phi +{\bf D})$$
$$Z_2:\qquad (\Theta +{\bf A},\Phi +{\bf B})$$
$$Z_3:\qquad (\Theta +{\bf A}+{\bf C},\Phi+{\bf B}+{\bf D})$$
The path integral measure therefore transforms by
\eqn\humphrey{\mu\to\mu\cdot (-1)^{
\sigma(\Theta +{\bf C},\Phi +{\bf D})+\sigma(\Theta +{\bf A},\Phi +{\bf B})
+\sigma(\Theta +{\bf A}+{\bf C},\Phi+{\bf B}+{\bf D})}.}
Using \sis\ and \ditor, this coincides with the desired transformation
law \mutu.\foot{In fact
a classic theorem (stated as Theorem 2 in \ref\atiyah{M. F. Atiyah,
``Riemann Surfaces And Spin Structures,'' Ann. Sci. Ecole Normale Sup.
{\bf 4} (1971) 47.}) asserts that
$$\sigma(\Theta+\ba+\bc,\Phi+\bb+\bd)=\sigma(\Theta,\Phi)
+\sigma(\Theta+\ba,\Phi+\bb)+\sigma(\Theta+\bc,\Phi+\bd)
+(\ba\cdot \bd-\bb\cdot \bc)$$
modulo 2.  This formula, which directly relates
the above expressions, is more or less equivalent to \sis, but
without need to choose a marking.}
This completes the proof in the path integral
formulation of how mirror symmetry acts for the ${\bf Z}_2\times {\bf Z}_2$
orbifold.

\subsec{Generalizations}

The above can be generalized in two ways: one can use a different
lattice with ${\bf Z}_2\times {\bf Z}_2$ symmetry, or one can
replace ${\bf Z}_2\times {\bf Z}_2$ with a different group.

For the first generalization, one can replace the hypercubic
lattice that we have used for simplicity with another lattice
that preserves the essential properties of the construction.
Those properties are the $\Gamma={\bf Z}_2\times {\bf Z}_2$ action
preserving the complex structure and holomorphic three-form, and
an additional operation
($Y_i\to -Y_i$ in the above) that reverses the complex structure
and commutes with ${\bf Z}_2\times {\bf Z}_2$.  The hypercubic
lattice is not the only one with these properties.  Some others
enter in \ref\fara{A. Faraggi, ``${\bf Z}_2\times {\bf Z}_2$ Orbifold
Compactification As The Origin Of Realistic Free Fermionic Models,''
Phys. Lett. {\bf B326} (1994) 62.}.  In some of these cases,
one obtains orbifolds with Hodge numbers
different from the one discussed above.

The other generalization involves replacing ${\bf Z}_2\times {\bf Z}_2$
with another group $\Gamma$.  We will briefly describe how this
can be done in substantial (but not complete) generality\foot{
All these considerations generalize to $n$-folds where
we replace $SO(3)$ with $SO(n)$.  In particular the discrete torsion
contains an element that comes from $H^2(SO(n),{\bf Z}_2)$, which
is related to how the twist elements lift from $SO(n)$ to $Spin(n)$.}.
Let $L$ be any three dimensional lattice in ${\bf R}^3$ and
$\Gamma\subset SO(3)$ a group of symmetries of $L$.
Let $S$ be the three dimensional torus $S={\bf R}^3/L$, and let
$S'$ be a second copy of $S$.
Consider the six dimensional torus
$T=S\times S'$.  Let $X_i$ and $Y_i$ be (local)
linear coordinates on $S$ and $S'$, respectively,
and pick the complex structure on $T$ such that $Z_i=X_i+iY_i$ are
local complex coordinates.  $T$ is a Calabi-Yau manifold with holomorphic
three-form $dZ_1\wedge dZ_2\wedge dZ_3$.
Consider the diagonal action of $\Gamma$
on $T$; that is, $\Gamma$ acts on both factors $S$ and $S'$.
Then $\Gamma$ preserves the complex structure and holomorphic three-form
of $T$, so we can consider the Calabi-Yau orbifold $T/\Gamma$.

$\Gamma$ also commutes with the operation $ Y_i\to -Y_i$
so as in the above discussion of ${\bf Z}_2\times {\bf Z}_2$,
an $R\to 1/R$ transformation on $S'$ can be used to show that the mirror
of $T/\Gamma$ is again $T/\Gamma$, with of course an inverted radius
of $S'$ and possibly a different discrete torsion.  As for ${\bf Z}_2
\times {\bf Z}_2$, the discrete torsion that is generated by the mirror
transformation can be computed by seeing how the path integral
measure transforms under $\eta_i\to -\eta_i$.   We will simply state
the results without proof.

A set of $\Gamma$ twists determines a flat $SO(3)$ bundle $E$,
via the embedding $\Gamma\subset SO(3)$.  For a given negative chirality
spin bundle
$\alpha$, the $\eta_i$ are a section of $\alpha\otimes E$.  The
path integral measure transforms as $(-1)^n$, where $n$ is the number modulo
2 of zero modes of the Dirac operator coupled to $\alpha\otimes E$.
That number is a topological invariant of $E$ regarded as an $SO(3)$
  bundle -- in computing it one can ignore the finer structure that $E$
has as a $\Gamma$-bundle. So the index
can only depend on $E$ through its one topological invariant,
which is the second Stieffel-Whitney class $w_2(E)\subset {\bf Z}_2$.
In fact, one can show (for instance, by deforming $E$ to a sum of line bundles)
that $n=\sigma_\alpha+w_2(E)$.  The transformation of the measure
under $\eta_i\to -\eta_i$ is thus
\eqn\hoko{\mu_{\alpha,E}\to (-1)^{\sigma_\alpha}(-1)^{w_2(E)}\mu_{\alpha,E}.}
As before, the factor of $(-1)^{\sigma_\alpha}$ means that this
is a mirror symmetry, and the factor $(-1)^{w_2(E)}$ means that the mirror
symmetry shifts the discrete torsion.  In fact, $(-1)^{w_2(E)}$
is the $E$ dependent phase factor associated with a particular
element $x\in H^2(\Gamma,U(1))$.  Under mirror symmetry, the discrete
torsion is multiplied by $x$; the $T/\Gamma $ theory with discrete torsion
$y$ is mirror to the same theory with discrete torsion $xy$.

One way to describe the torsion element $x\in H^2(\Gamma,U(1))$ more
explicitly is as follows.  We will describe an element $\widehat x\in
H^2(SO(3),U(1))$ which, for any $\Gamma\subset SO(3)$, restricts to the
required $x$. $H^2(SO(3),U(1))$ classifies extensions of $SO(3)$ by $U(1)$.
One such extension is the group $U(2)$; that is, the center of $U(2)$
is isomorphic to $U(1)$ and the quotient $U(2)/U(1)$ is isomorphic to
$SO(3)$:
\eqn\jujj{1\to U(1)\to U(2) \to SO(3)\to 1.}
The element of $H^2(SO(3),U(1))$ associated with this extension
is the desired $\widehat x$.

\newsec{A ${\bf Z}_3\times {\bf Z}_3$ Example}

In this section, we will describe a ${\bf Z}_3\times {\bf
Z}_3$
orbifold that is somewhat similar to the ${\bf Z}_2\times {\bf Z}_2$
example.  To begin with, we need a genus one   curve $E$ with a ${\bf Z}_3$
symmetry that has non-trivial fixed points.  This curve can be regarded
as the complex $z$ plane divided by a hexagonal lattice (its complex
structure is uniquely determined since the hexagonal lattice is the only
lattice in the plane with ${\bf Z}_3$ symmetry).  The ${\bf Z}_3$
symmetry is generated  by $z\to \zeta z$, with $\zeta=\exp(2\pi i/3)$.

Alternatively, the same curve can be described algebraically
 by the equation in homogeneous variables
\eqn\theeq{y^3=u^3+v^3.}
The ${\bf Z}_3$ symmetry is then generated by $y\to \zeta y$ (with $u,v$
invariant).  The ${\bf Z}_3$ action on $E$ has three fixed points,
the points with $y=0$, $u^3+v^3=0$.

Now we introduce three identical curves $E_i$, $i=1\dots 3$; $E_i$
is the quotient of the $z_i$ plane by a hexagonal lattice or alternatively
is given by equations
\eqn\thereq{y_i{}^3=u_i{}^3+v_i{}^3}
in homogeneous variables $y_i,u_i,v_i$.
On $T=E_1\times E_2\times E_3$, there is a natural action of $\Gamma_0=
{\bf Z}_3\times
{\bf Z}_3\times {\bf Z}_3$.
The subgroup of $\Gamma_0$ that preserves the holomorphic three-form
of $T$ (which is $\omega=dz_1\wedge dz_2\wedge dz_3$) is
the group of transformations $z_i\to \zeta^{a_i}z_i$ with
$\zeta^{a_1+a_2+a_3}=1$.  We call this group $\Gamma$; it is isomorphic to
${\bf Z}_3\times {\bf Z}_3$.  We wish to study the orbifold $T/\Gamma$.

The possible
discrete torsion in this theory
 can be described very explicitly.  If $T_\sigma$
and $T_\tau$ are two elements of $\Gamma$, say $T_\sigma=(\zeta^{a_1},
\zeta^{a_2},\zeta^{a_3})$ and $T_\tau=(\zeta^{b_1},
\zeta^{b_2},\zeta^{b_3})$, then the torsion factor
in \ritet\ is
\eqn\torfac{\epsilon(T_\sigma,T_\tau)=\zeta^{m\sum_{i,j,k=1}^3\epsilon_{ijk}
a_jb_k}}
where $\epsilon_{ijk}$ is the antisymmetric tensor with $\epsilon_{123}=+1$.

The classical geometry can be studied similarly to the ${\bf Z}_2\times
{\bf Z}_2$ example.  The group element $\alpha=(\zeta,\zeta^{-1},1)$
has a fixed point set consisting of nine tori.
Allowing also for fixed tori of other group elements, there are $3\times 9=
27$ fixed tori in all.
  In addition, there are $3\times 3\times 3=27$
fixed points of the whole group; as in the ${\bf Z}_2\times {\bf Z}_2$
example, these are intersections of three fixed tori.

A difference from the
${\bf Z}_2\times {\bf Z}_2$ example is that in addition to the identity
element and group elements that have fixed tori,
${\bf Z}_3\times {\bf Z}_3$ also contains the elements $\beta=(\zeta,\zeta,
\zeta)$ and $\beta^2$ that act with isolated
fixed points (27 of them) rather than
fixed tori.

\subsec{Spectrum Of The Orbifold}

Now let us determine the Ramond ground states of the
orbifold.  First we work in the absence of discrete torsion.

Each of the $E_i$ has the familiar Hodge diamond
\eqn\pomm{\pmatrix{1 & 1\cr
                   1 & 1\cr}.}
In the ${\bf Z}_3$ action on the cohomology, $H^0$ and $H^2$ are invariant
but $H^{1,0}$ (which is generated by $dz_i$) and $H^{0,1}$ (which is
generated by $d\bar z_i$) transform with eigenvalue $\zeta$ or $\zeta^{-1}$.
The $\Gamma$-invariant part of the cohomology of $E_1\times E_2\times E_3$
can be represented by the Hodge diamond
\eqn\qomm{\pmatrix{ 1 & 0 & 0 & 1\cr
                    0 & 0 & 3 & 0\cr
                    0 & 3 & 0 & 0\cr
                    1 & 0 & 0 & 1\cr}.}
This is the contribution of the untwisted sector to the cohomology of the
orbifold.

Next we consider the sector twisted by $\alpha$.  The low-lying states
are obtained by quantizing the fixed point set, which as we noted
above consists of nine fixed tori.
The spectrum of R ground states in the twisted Hilbert space ${\cal H}_\alpha$
consists therefore of nine copies of the cohomology of a torus;
with the shift in the zero point values of the $U(1)$ charges,
$H^{p,q}$ of the torus contributes to $H^{p+1,q+1}$ of the orbifold.
The contribution of the nine tori
can hence be represented by the Hodge diamond
\eqn\bmattto{\pmatrix{ 0 & 0 & 0 & 0 \cr
                       0 & 9 & 9 & 0 \cr
                       0 & 9 & 9 & 0 \cr
                       0 & 0 & 0 & 0 \cr}.}
Now we must project onto the $\Gamma$-invariant part of the cohomology;
in fact, the $\Gamma$-invariant states are those that contribute
 to $H^{1,1}$ and $H^{2,2}$ (coming from
$H^{0,0}$ and $H^{1,1}$ of the torus).  As there are altogether six
group elements similar to $\alpha$, the contribution to the Hodge diamond
from this source is
\eqn\nmatto{\pmatrix{ 0 & 0 & 0 & 0 \cr
                       0 & 0 & 54 & 0 \cr
                       0 & 54 & 0 & 0 \cr
                       0 & 0 & 0 & 0 \cr}.}

We also have the sectors twisted by $\beta$ or $\beta^2$.
The fixed point set consists in each case of 27 isolated points.
The cohomology of a point consists of the one-dimensional space
$H^{0,0}$ with trivial $\Gamma$ action.  But the shift in the zero
point of the $U(1)$ charges shifts the contribution of the $\beta$ twisted
sector to $H^{1,1}$ and that of the $\beta^2$ twisted sector to
$H^{2,2}$.  So the contribution
to the Hodge diamond is
\eqn\inmatto{\pmatrix{ 0 & 0 & 0 & 0 \cr
                       0 & 0 & 27 & 0 \cr
                       0 & 27 & 0 & 0 \cr
                       0 & 0 & 0 & 0 \cr}.}
Adding it all up, the Hodge diamond of the orbifold without discrete
torsion is
\eqn\enmatto{\pmatrix{ 1 & 0 & 0 & 1 \cr
                       0 & 0 & 84 & 0 \cr
                       0 & 84 & 0 & 0 \cr
                       1 & 0 & 0 & 1 \cr}.}
In particular, there are 84 Kahler deformations and no complex structure
deformations.

\bigskip
\noindent{\it Inclusion Of Discrete Torsion}

Now we repeat the analysis in the presence of discrete torsion.
We may as well pick $m=1$ in \torfac, since the $m=2$ case differs
by a permutation of the $E_i$.

The contribution to the space of
R ground states from the untwisted sector is unaffected by discrete
torsion since $\epsilon(T_\sigma,T_\tau)=1$ if $T_\sigma=1$.

Now let us consider the $\alpha$-twisted sector.  Though
 $\epsilon(\alpha,\alpha)
=1$, we have $\epsilon(\alpha,\tilde\alpha)= \zeta^{-1}$ where $\tilde\alpha
=(1,\zeta,\zeta^{-1})$.  Hence in the presence of discrete torsion,
projecting onto the $\Gamma$-invariants means projecting onto states
that in the natural action of $\Gamma$ transform under $\tilde\alpha$ as
$\zeta$.  As $\tilde \alpha$ acts on $E_3$ by $z_3\to \zeta^{-1}z_3$,
the only state in the cohomology of $E_3$ that transforms as $\zeta$
in the natural action of $\tilde\alpha$ on the cohomology of $E_3$
is $d\bar z_3$, which generates $H^{0,1}$.  Therefore (upon allowing
for the shift in the zero point) the nine fixed
tori of $\alpha$, which are copies of $E_3$, contribute to $H^{1,2}$
of the orbifold.  In all, of the six ${\bf Z}_3\times {\bf Z}_3$ elements
obtained from $\alpha$ by permutations of the $E_i$, three contribute
to $H^{1,2}$ of the orbifold and three to $H^{2,1}$.  The total contribution
from these sectors is therefore
\eqn\pumatrix{\pmatrix{ 0 & 0 & 0 & 0 \cr
                        0 & 27 & 0 & 0 \cr
                        0 & 0 & 27 & 0 \cr
                        0 & 0 & 0 & 0 \cr}.}

Similarly, with discrete torsion, to get the contribution of the
$\beta$-twisted
sector we must project onto  the part of the cohomology
of the fixed point set of $\beta$ that transforms as $\zeta$ under $\alpha$.
But the fixed point set of $\beta$ is a set of 27 points, which are
all left fixed by $\alpha$ so that $\alpha$ acts trivially on their cohomology;
hence, with discrete torsion,
the $\beta$-twisted sector does not contribute any R ground states.
The same goes for $\beta^2$.

The overall Hodge diamond of the theory with discrete torsion is therefore
\eqn\pumatrix{\pmatrix{ 1 & 0 & 0 & 1 \cr
                        0 & 27 & 3 & 0 \cr
                        0 & 3 & 27 & 0 \cr
                        1 & 0 & 0 & 1 \cr}.}
This spectrum is not mirror to \enmatto, which should come as no surprise
since the construction described in section 3 does not apply.

\subsec{Comparison To Classical Geometry}

Since the conformal field theory
without discrete torsion has Kahler deformations and
no complex structure deformations, it should be compared (as in \aspin)
to the blow-up of the orbifold.  On the other hand, with discrete torsion
there are complex structure modes, and one wonders to what extent the conformal
field theory with discrete torsion
can be compared to the conformal field theory of a smooth Calabi-Yau manifold
obtained by deforming $T/\Gamma$.

Let us find a  family of smooth Calabi-Yau manifolds
to which the orbifold $T/\Gamma$ can be deformed.
To this aim, we want an algebraic description of the orbifold.
Beginning with the algebraic description \thereq\ of the torus $T$,
we simply project onto the $\Gamma$-invariant polynomials.  They
are $u_i, $ $v_i$, and $y=y_1y_2y_3$  ($y$ is of degree one
under scaling of any pair $u_i,v_i$), obeying the single equation
\eqn\hikki{y^3=\prod_{i=1}^3(u_i{}^3+v_i{}^3).}
This can be deformed to
\eqn\ikki{y^3=F(u_i,v_i)+yG(u_i,v_i),}
with $F$ a function that is cubic in each pair of variables $u_i,v_i$ and
$G$  quadratic in each pair.  (This preserves the homogeneity of the
equation.  Note that a term $y^2H(u_i,v_i)$ need
not be included as it could
be eliminated by $y\to y+H/3$.)
For generic $F,G$, we get a smooth Calabi-Yau manifold.

Let us count the parameters in \ikki.  The space of quartic polynomials
in $u_i,v_i$ is of dimension 4.  The space of cubic polynomials is of dimension
3.  So the space of $F$'s and $G$'s is of dimension $4^3+3^3=91$.
After removing one for an overall scaling and $3\times 3=9$ to allow
for the $SL(2,{\bf C})$ action on $u_i,v_i$, we are left with $81$ parameters
in the equations.   These are the right parameters since
in the present example
the polynomial deformations can be shown to faithfully represent the possible
deformations of the complex structure.

\subsec{Origin Of The Discrepancy}

So once again we have a discrepancy: the number of complex structure
deformations in the conformal field theory of the orbifold (with discrete
torsion) is 27, but in the classical geometry there are 81.
The number of missing modes is 54, which equals $2\times 27$, where
27 is the number of ${\bf Z}_3\times {\bf Z}_3$ fixed points.\foot{There
is no such discrepancy for the Kahler deformations.  One can show -- for
instance by computing the Euler characteristic -- that the smooth Calabi-Yau
manifold given by a generic equation \ikki\ has $b^{1,1}=3$, in agreement
with the conformal field theory with discrete torsion.}

So, as in the ${\bf Z}_2\times {\bf Z}_2$
 example, one can suspect that the singularities
at the fixed
points are not completely eliminated by the complex structure deformations,
and that one instead gets 27 singularities each of which has 2 relevant
operators.  As we explained earlier,
the only isolated singularity with precisely 2 relevant
operators is
\eqn\hurrof{y^3=u_1{}^2+u_2{}^2+u_3{}^2,}
and if one wishes to have 27 singularities with this structure, one would
have to impose $2\times 27= 54$
conditions on the parameters in \ikki.   The proposal
is then that the orbifold with discrete torsion (and generic complex
structure deformation) corresponds to a specialization of \ikki\
with 54 conditions imposed to ensure 27 singularities of this type.

As in the ${\bf Z}_2\times {\bf Z}_2$ case, evidence for this interpretation
can be found by considering in more detail the possible perturbations
of \ikki.  The space $V_i$
of cubic polynomials in $u_i,v_i$ is four-dimensional.
Pick in this space a three-dimensional complement $W_i$ to the one-dimensional
subspace generated by $u_i{}^3+v_i{}^3$. Let $F_i=u_i{}^3+v_i{}^3$,
and let $\delta F_i$ be an element of $W_i$.

Then -- apart from modes that can be eliminated by scaling and $SL(2,{\bf C})$
-- the modes in \ikki\  can be written in detail as follows.
There are $3\times 3\times 3=27$ modes in which to the unperturbed
equation \hikki\ one adds
\eqn\humph{\delta F_1\delta F_2 F_3+F_1\delta F_2\delta F_3+\delta F_1 F_2
\delta F_3.}
There are $27$ more modes of the form
\eqn\gumph{\delta F_1\delta F_2\delta F_3}
and 27 of the
form
\eqn\tumpth{yG(u_i,v_i).}

The 27 complex structure modes of the conformal field theory with
discrete torsion each arose as a contribution of a particular fixed
torus from a particular twisted sector.    The mode associated with
a given fixed torus must deform the singularity of that torus but not
the others.  The modes that do this are the ones in \humph.
For instance, any torus fixed by  $\alpha=(\zeta,\zeta^{-1},1)$
 lies at $F_1=F_2=0$, so the
associated modes should not vanish if $F_1=F_2=0$, but should vanish
if $F_1=F_3=0 $ or $F_2=F_3=0$ (to avoid deforming other fixed tori).
The modes in the $\alpha$-twisted sector are
$\delta F_1\delta F_2F_3$.
Similarly, the other twisted sectors give the other modes in \humph.
These modes all vanish at the fixed points $F_1=F_2=F_3$
and do not have the flexibility to eliminate the singularities at the fixed
points.  The modes in \gumph\ and \tumpth\ would do this, but
do not arise in the conformal field theory.

The 27 surviving singularities
are of the form \hurrof\ as this is the only isolated singularity
with two relevant operators.
As a check, let us note that in \hurrof\ there is a ${\bf Z}_3$ symmetry
$y\to \zeta y$ (with the $u_i$ fixed); the relevant operators $1$ and $y$
transform as $1$ and $\zeta$.  Similarly, the orbifold theory \hikki\
has a ${\bf Z}_3$ symmetry $y\to \zeta y$; the 54 missing perturbations
include 27 modes \gumph\  that are invariant and 27 modes
\tumpth\ transforming as $\zeta$. This is in agreement with what one
would expect for 27 singularities of the structure claimed.

\bigskip
\noindent{\it The Support Of The Torsion}

For the same reasons as in the ${\bf Z}_2\times {\bf Z}_2$ case,
the discrete torsion is non-zero in an arbitrarily small neighborhood
of the singularity in \hurrof.  On the other hand, the complement $W$
of the singularity has the cohomology of ${\bf S}^5$
\ref\milnor{J. Milnor, {\it Singular Points Of Complex Hypersurfaces,}
(Princeton University Press, 1968), p. 72.}.  In particular,
$H^2(W,U(1))$ vanishes and with it the discrete torsion.
The effects of the underlying discrete torsion
are thus supported at the origin.

In fact, from \milnor, one can make a stronger statement:
the singular space \hurrof\ is topologically equivalent to ${\bf R}^6$,
and thus we have found a classical solution of string theory that
can be interpreted as a kind of stringy ``lump'' in ${\bf R}^6$,
whose complement is isomorphic topologically to the complement of
an ordinary ball in ${\bf R}^6$.  However, the (singular) Calabi-Yau
metric on the space \hurrof\ is presumably not asymptotic at infinity
to the flat metric on ${\bf R}^6$.
\bigskip
We would like to thank A. Faraggi and K.S. Narain for discussions
on ${\bf Z}_2\times {\bf Z}_2$ orbifolds and P. Aspinwall,
T. Hubsch, S. Katz and D. Morrison
for useful advice about Calabi-Yau manifolds.

The research of C.V. was supported in part by the Packard fellowship and NSF
grants PHY-92-18167 and PHY-89-57162.  The work of E.W. was supported
in part by NSF Grant PHY92-45317.
\listrefs
\end